\documentclass[showpacs,aps,graphicx,twocolumn]{revtex4}
\usepackage{graphicx}
\usepackage{verbatim}
\usepackage{algorithm}
\usepackage{algorithmic}
\usepackage{subfigure}
\usepackage[caption=false,font=scriptsize,labelfont=sf,textfont=sf]{subfig}

\begin{document}
\title{Gradient-Free optimization algorithm for single-qubit quantum classifier}

\author{Anqi Zhang$^{1}$, Xiaoyun He$^{1}$, Shengmei Zhao$^{1,2}$\footnote{Email address:
zhaosm@njupt.edu.cn}}
\address{
$^1$Institute of Signal Processing Transmission, Nanjing University of Posts and Telecommunications (NUPT), Nanjing, 210003,  China \\
$^2$Key Lab of Broadband Wireless Communication and Sensor Network Technology, Ministry of Education, Nanjing, 210003, China\\
}
\date{\today }

\begin{abstract}
	In the paper, a gradient-free optimization algorithm for single-qubit quantum classifier is proposed to  overcome the effects of barren plateau caused by quantum devices. A rotation gate $R_{X}(\phi)$ is applied on a single-qubit binary quantum classifier, and the training data and parameters are loaded into $\phi$ with the form of vector-multiplication. The cost function is decreased by finding the value of each parameter that yield the minimum expectation value of measuring the quantum circuit. The algorithm is performed iteratively for all parameters one by one, until the cost function satisfies the stop condition. The proposed algorithm is demonstrated for a classification task and is compared with that using Adam optimizer. Furthermore, the performance of the single-qubit quantum classifier with the proposed gradient-free optimization algorithm is discussed when the rotation gate in quantum device is under different noise. The simulation results show that the single-qubit quantum classifier with proposed gradient-free optimization algorithm can reach a high accuracy faster than that using Adam optimizer. Moreover, the proposed gradient-free optimization algorithm can quickly completes the training process of the single-qubit classifier. Additionally, the single-qubit quantum classifier with proposed gradient-free optimization algorithm  has a good performance in noisy environments.
\end{abstract}

\maketitle
\section{Introduction}

Classification has been one of the main issues in Quantum Machine Learning\cite{JBiamonte,MBenedetti,SLTeresa,KMNaoko}. There are several types of quantum classifiers, some are inspired by their classical counterparts with their kernel parts replaced by a high-depth quantum circuit\cite{KHWan,ETorrontegui}, some are inspired by neural networks with multi-layer quantum cuicuit 
structures\cite{JMArrazola,AMari,NKilloran,AGilyn,ECampos}, and others are termed quantum circuit learning with a hybrid quantum-classical (HQC) structure\cite{NWiebe,SDangwal,RKune,AChalumuri,ASBhatia,APerezSalinas,SAdhikary,AQZhangMulti}. 

By using a small amount of quantum resources, quantum classifiers based on HQC structure have become a promising candidate in the Noisy Intermediate Scale Quantum(NISQ) era. For examples, Ref.\cite{APerezSalinas} presented a multiple data re-uploading method to increase circuit express ability, in which a quantum circuit was organized as a series of data re-uploading and single-qubit processing units. Ref.\cite{SAdhikary} built a model with only two qubits in quantum circuit that could simultaneously manipulate two classes of training samples. Furthermore, Ref.\cite{AQZhangMulti} trained multi-class data, more than 2 classes, simultaneously for classification tasks in the qubit system.

The gradient-based parameter optimization method inside quantum classifier, such as Adam optimizer, may cause the barren plateau\cite{ZHolmes} (i.e., vanishing gradient) problem. There are some solutions to overcome this issue. For instance, a layer wise learning optimization method was proposed by A Skolik \emph{et al.}\cite{ASkolik} to update only subset of parameters in each training iteration. A Bayesian optimization(BO) based on Gaussian process regression(GPR) and noisy expected improvement(NEI) was proposed by G Iannelli \emph{et al.}\cite{GIannelli} to provide a more precise estimation of the ground state energy in a few iterations. An efficient method for simultaneously optimizing both the structure and parameter values of quantum circuits was proposed by M Ostaszewski \emph{et al.}\cite{MOstaszewski} by finding the angle of rotation that yield the minimum cost value directly for all the parameterized gates. However, these works focus mainly on applying the methods in the variational quantum eigensolver(VQE) area, it may not yield a higher accuracy when transfer these solutions to solve the classification tasks directly.

In this work, we propose a gradient-free optimization algorithm for a single-qubit quantum classifier.
In this algorithm, a rotation gate $R_{X}(\phi)$ applied on one qubit initialized as $|0\rangle$ can be regard as a binary quantum classifier. Training data and parameters of the quantum classifier are loaded into $\phi$ with the form of vector-multiplication. The cost function is decreased by finding the value of each parameter that yield the minimum expectation value of measuring the quantum circuit until it meet the stop condition. After several iterations, the cost function is converged and the optimal parameters are obtained. The classification information of an unclassified data can be determined by the result with the highest probability after measuring the circuit.

The advantages of the proposed gradient-free optimization algorithm are three-fold:
1)Compared with gradient-based algorithm, gradient-free optimization algorithm requires lesser computations, which can save more computing resources; 2)Parameters are loaded as weights of each component of training data in single-qubit quantum classifier, which can save more qubits and quantum gates resources and further reduce the barren plateau problem; 3)The updated value of parameter is calculated with the usage of statistical average method, which can increase the accuracy of classification.

\section{Gradient-Free optimization algorithm for single-qubit quantum classifier}
In this section, we present the proposed gradient-free optimization(GFO) algorithm for a single-qubit quantum classifier in details. 

\begin{figure*}[t]
	\centering
	\includegraphics[width=0.7\textwidth]{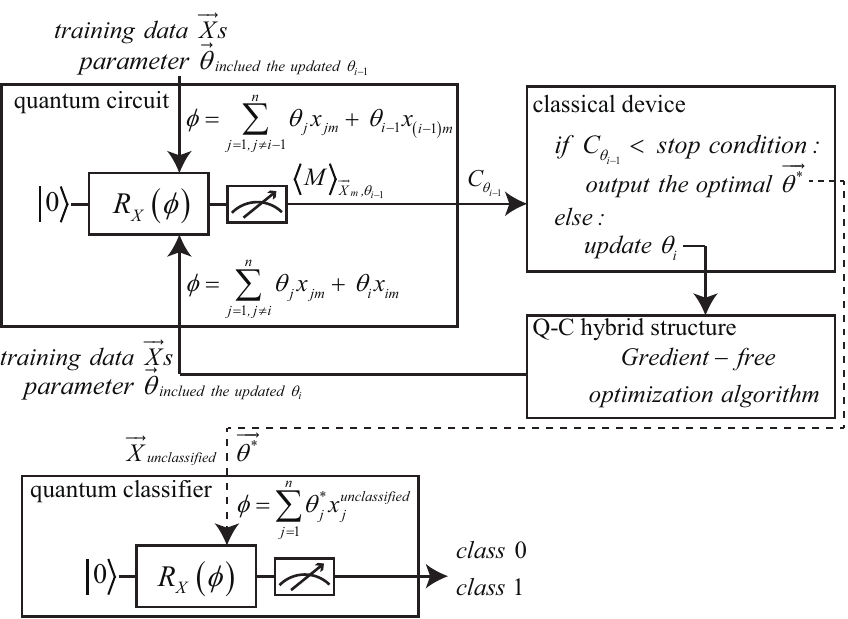}
	\caption{The framework for optimizing parameters of the single-qubit quantum classifier. Arrow lines in the figure represent the flow of information. In the training processes (body lines), training data $\vec{X}$s and parameters $\vec{\theta}$ are fed into the position of $\phi$ of the rotation gate with the form of vector-multiplication. The value of cost function is obtained by measuring the quantum circuit and check that whether it's lesser than the given stop condition. If not, continue the next parameter-optimizing iteration where other parameters are fixed but update the $\theta_{i}$ only; if so, turn into the testing processes (dot lines), feed the optimal parameters vector $\vec{\theta}^{*}$ and an unclassified data $\vec{X}_{unclassified}$ into the same circuit which named single-qubit quantum classifier. The classification information of $\vec{X}_{unclassified}$ can be specified by the result with the highest probability after measuring the circuit.}
	\label{Fig:1}
\end{figure*}

The framework for optimizing parameters of the single-qubit quantum classifier is given in Fig.1. Due to the nature of the GFO algorithm that only one of all parameters in quantum circuit is updated in each iteration while other parameters are fixed, we suppose that the former $i-1$ parameters are have been updated already. In training processes, a rotation gate $R_X(\phi) = exp(-i\frac{\phi}{2}\sigma_{X})$ applying on one qubit is used as the quantum circuit, where $\phi\in(-\pi,\pi]$ is the angle of rotation and $\sigma_{X}$ is the Pauli-X matrix. $\vec{X}=(x_1,\ldots,x_n)$ is a $n$-dimension training data in a batch used for updating the ($i-1$)-th parameter and $\vec{X}$s are selected randomly from the training dataset which is defined as $S=\{\vec{X},f(\vec{X})\}$, $f(\vec{X})=$0 or 1 is the label of $\vec{X}$. 
The method of loading training data and parameters is to perform multiplication operation between two $n$-dimensional vectors, $\vec{X}$ and $\vec{\theta}$, which makes two inputs into a single number, e.g. $\phi=\sum_{i=1}^{n}\theta_{i}x_{i}$, $\vec{\theta}=(\theta_1,\ldots,\theta_n)$ is the parameter vector need to be tuned for the quantum classifier. After loading a batch of training data $\vec{X}$s and parameter vector $\vec{\theta}$ into the quantum circuit, the value of cost function $C_{\theta_{i-1}}$ specified by the ($i-1$)-th parameter, where $C_{\theta_{i-1}}=\frac{1}{M}\sum_{m=1}^{M}(1-|\langle M\rangle_{\vec{X}_{m},\theta_{i-1}}|^{2})$, is obtained by using Hermitian operator $M$ to measure the quantum circuit. $\vec{X_{m}}$ is the $m$-th training data in the batch. 
$C_{\theta_{i-1}}$ is checked whether it satisfy the stop condition of training processes. If so, output the optimal parameter vector $\vec{\theta}^{*}$ for testing processes; if not, continue to update the $i$-th parameter $\theta_{i}$ in $\vec{\theta}$. In testing processes, an unclassified data $\vec{X}_{unclassified}$ and $\vec{\theta}^{*}$ are loaded into the same quantum circuit termed single-qubit quantum classifier. The classification information of $\vec{X}_{unclassified}$ can be obtained after performing measurements. 

Here, we focus on the details of finding the optimal parameter vector $\vec{\theta}^{*}$. 

Inspired by optimization strategies in Ref.\cite{MOstaszewski}, the expectation value of a quantum circuit with structure $U$ and initial state $\rho$ can be rewritten as a function of angle $\phi$, which has sinusoidal form as $\langle M\rangle_{\phi}=Tr(MU\rho U^{\dag})=Asin(\phi+B)+C$, where A, B and C are coefficients need to be determined. In this work, we consider the task of minimizing the value of cost function $C_{\theta_{i}}$ by maximizing every $|\langle M\rangle_{\vec{X}_{m},\theta_{i}}|$ part. Clearly, it can be ensured if the coefficients are estimated. Therefore, the expectation value characterized by the $i$-th parameter when training data is $\vec{X_{m}}$ can be expressed as

\begin{eqnarray}\label{eqn1}
	&&\langle M\rangle_{\theta_{im}} \nonumber \\
	&=& Tr[M R_{X}(\phi)\rho R_{X}^{\dag}(\phi)] \\ \nonumber
	&=& Tr[M e^{-i\sigma_{X}(\theta_{im}x_{im} + \vec{\theta}_{m}^{'}\vec{X}_{m}^{'})/2}\rho e^{i\sigma_{X}(\theta_{im}x_{im} + \vec{\theta}_{m}^{'}\vec{X}_{m}^{'})/2}] \\ \nonumber
	&=&  \frac{1}{2}[(\langle M\rangle_{0}-\langle M\rangle_{\pi})cos(\vec{\theta}_{m}^{'}\vec{X}_{m}^{'})+ \\ \nonumber
	& &(\langle M\rangle_{\frac{\pi}{2}}-\langle M\rangle_{-\frac{\pi}{2}})sin(\vec{\theta}_{m}^{'}\vec{X}_{m}^{'})]cos(\theta_{im}x_{im})  \\\nonumber
	&+& \frac{1}{2}[(\langle M\rangle_{\frac{\pi}{2}}-\langle M\rangle_{-\frac{\pi}{2}})cos(\vec{\theta}_{m}^{'}\vec{X}_{m}^{'})+ \\\nonumber
	& &(\langle M\rangle_{0}-\langle M\rangle_{\pi})sin(\vec{\theta}_{m}^{'}\vec{X}_{m}^{'})]sin(\theta_{im}x_{im})  \\ \nonumber
	&+&\frac{1}{2}(\langle M\rangle_{0}+\langle M\rangle_{\pi}), \\ \nonumber
\end{eqnarray}
where $\vec{\theta}_{m}^{'}\vec{X}_{m}^{'} = \sum_{j=1}^{n}\theta_{jm}x_{jm},j\ne i$, and other $n-1$ parameters, except $\theta_{i}$, are fixed to their current values. $\langle M\rangle_{0}$, $\langle M\rangle_{\pi}$, $\langle M\rangle_{\pi/2}$ and $\langle M\rangle_{-\pi/2}$ are achieved by set the value of $i$-th parameter as 0, $\pi$, $\pi/2$, $-\pi/2$. From the sine function of the identity $acos(x)+bsin(x)+c= \sqrt{a^{2}+b^{2}}sin(x+arctan2(a,b))+c$, we can obtain a compact expression of Eq.(1), which it's easy to locate the maxima at $\theta_{im}x_{im}=\pi/2$. Therefore, the updated value of the $i$-th parameter has the expression as

\begin{eqnarray}\label{eqn3}
	&&\theta_{im}^{upd}  \\ \nonumber
	&=& \mathop{\arg\max}_{\theta_{im}}\langle M\rangle_{\theta_{im}}\\ \nonumber
	&=& \frac{1}{x_{im}}\frac{\pi}{2}-\frac{1}{x_{im}}arctan2[(\langle M\rangle_{0}-\langle M\rangle_{\pi})cos(\vec{\theta}_{m}^{'}\vec{X}_{m}^{'})+ \\ \nonumber
	& &(\langle M\rangle_{\frac{\pi}{2}}-\langle M\rangle_{-\frac{\pi}{2}})sin(\vec{\theta}_{m}^{'}\vec{X}_{m}^{'}), \\ \nonumber
	& &(\langle M\rangle_{\frac{\pi}{2}}-\langle M\rangle_{-\frac{\pi}{2}})cos(\vec{\theta}_{m}^{'}\vec{X}_{m}^{'})+ \\ \nonumber
	& &(\langle M\rangle_{0}-\langle M\rangle_{\pi})sin(\vec{\theta}_{m}^{'}\vec{X}_{m}^{'})].
\end{eqnarray}

The $i$-th parameter can not be optimized by one training data but a batch, a strategy of make the average value of $m$ $i$-th updated parameters is used in each iteration, 
\begin{eqnarray}\label{eqn4}
	\theta_i^{upd} = \frac{1}{M}\sum_{m=1}^{M}\theta_{im}^{upd},
\end{eqnarray}
where $M$ is the number of training date in a batch. The algorithm finds each updated parameter in $\vec{\theta}$ sequentially for all $i=1,\ldots,n$ in order to complete cycle by cycle until a stop condition is met. The gradient-free method is summarized in Algorithm 1.

\begin{algorithm}
	\caption{Gradient-free optimization algorithm for single-qubit quantum classifier}
	\label{alg1} 
	\begin{algorithmic}[1] 
		\REQUIRE Initialize the parameters $\theta_i$ randomly, where $1\leq i\leq n$; set batch size $M$(we set $M$ as 10 in this work); a stop condition; 
		\LOOP 
		\STATE Calculate the value of cost function as cost-now.
		\FOR{$i=1$ to $n$} 
		\STATE For all parameters except the $i$-th one.
		\FOR{$m=1$ to $M$}		
		\STATE {Estimate $\langle M\rangle_0,\langle M\rangle_{\pi},\langle M\rangle_{\frac{\pi}{2}},\langle M\rangle_{-\frac{\pi}{2}},$
		
		and calculate $\vec{\theta}_{m}^{'}\vec{X}_{m}^{'};$
		
		$a=\frac{1}{2}[(\langle M\rangle_0-\langle M\rangle_{\pi})cos(\vec{\theta}_{m}^{'}\vec{X}_{m}^{'})+$
		$(\langle M\rangle_{\frac{\pi}{2}}-\langle M\rangle_{-\frac{\pi}{2}})sin(\vec{\theta}_{m}^{'}\vec{X}_{m}^{'})];$
		
		$b=\frac{1}{2}[(\langle M\rangle_{\frac{\pi}{2}}-\langle M\rangle_{-\frac{\pi}{2}})cos(\vec{\theta}_{m}^{'}\vec{X}_{m}^{'})+
		(\langle M\rangle_0-\langle M\rangle_{\pi})sin(\vec{\theta}_{m}^{'}\vec{X}_{m}^{'})];$
		
		$\theta_{im}^{opt} \leftarrow \frac{1}{X_{im}}(\frac{\pi}{2}-arctan2(a,b)).$} 
		\ENDFOR
		
		Get the average: $\theta_i^{opt} = \frac{1}{M}\sum_{m=1}^{M}\theta_{im}^{opt}$. 
		
		Calculate cost value: cost-new.
		
		\IF{cost-new $<$ cost-now} 
		\STATE $\theta_i^{opt}$ changed; 
		
		cost-now $=$ cost-new.
		\ELSE 
		\STATE $\theta_i^{opt}$ remain the same. 
		\ENDIF 	 
		\ENDFOR	{ if the condition is met.}
		\ENDLOOP	
	\end{algorithmic} 
\end{algorithm}

\section{Simulation}
In this section, we demonstrate the feasibility of the proposed GFO algorithm by simulations. Here, we use 10000 training samples and 1000 testing samples for each class, which are selected randomly from the MNIST dataset. We deal with each sample into 32 dimensions by Rough Grid Feature method\cite{PaoloC}, and use PennyLane\cite{VilleB} module in Python to realize the proposed algorithm. The simulations of GFO algorithm are performed separately based on circuits with and without noise deriving from the imperfection of quantum gate. We compare the optimizing results of the proposed GFO algorithm with Adam optimizer. Parameters of the quantum circuit has been updated for 15 loops each simulation.

\begin{figure*}[htbp]
	\centering
	\subfigure[] {\includegraphics[width=.45\textwidth]{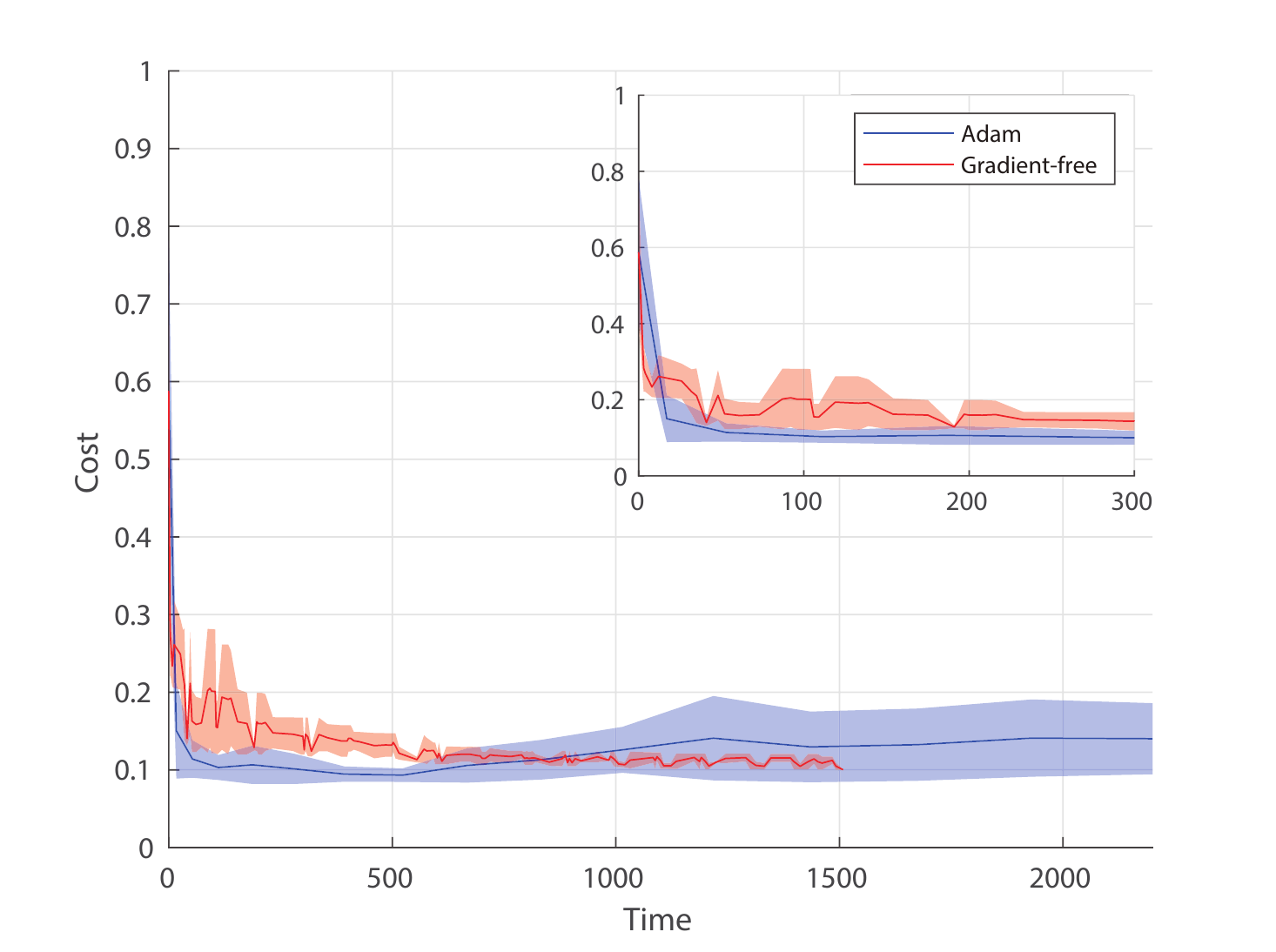}}
	\subfigure[] {\includegraphics[width=.45\textwidth]{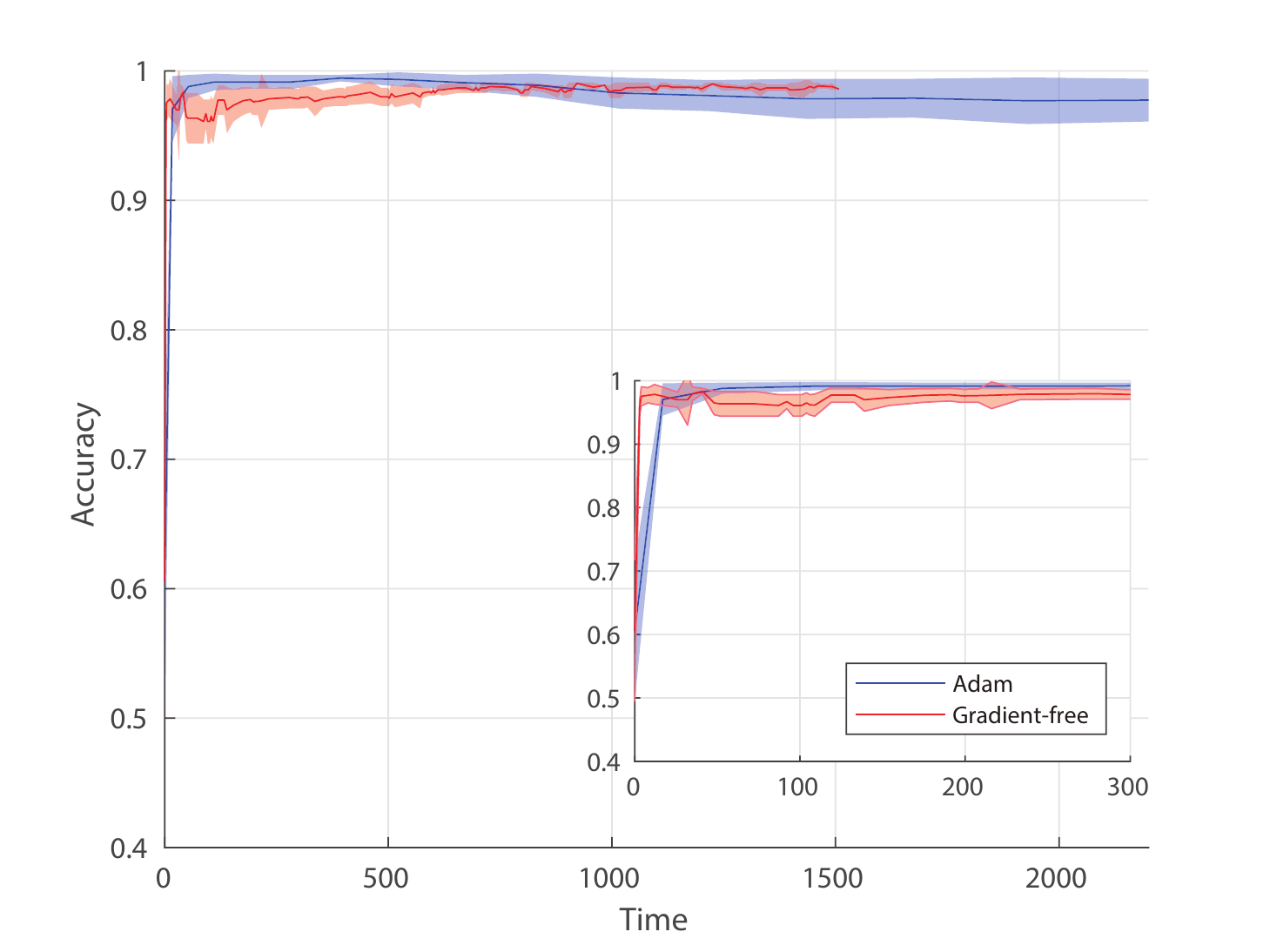}}
	\caption{The classification accuracy and the cost function against the time of 15 loops for the two-class classification task, where (a) is the classification accuracy and (b) is the cost function.}
	\label{Fig:2}
\end{figure*}

\begin{figure*}[htbp]
	\centering
	\subfigure[] {\includegraphics[width=.45\textwidth]{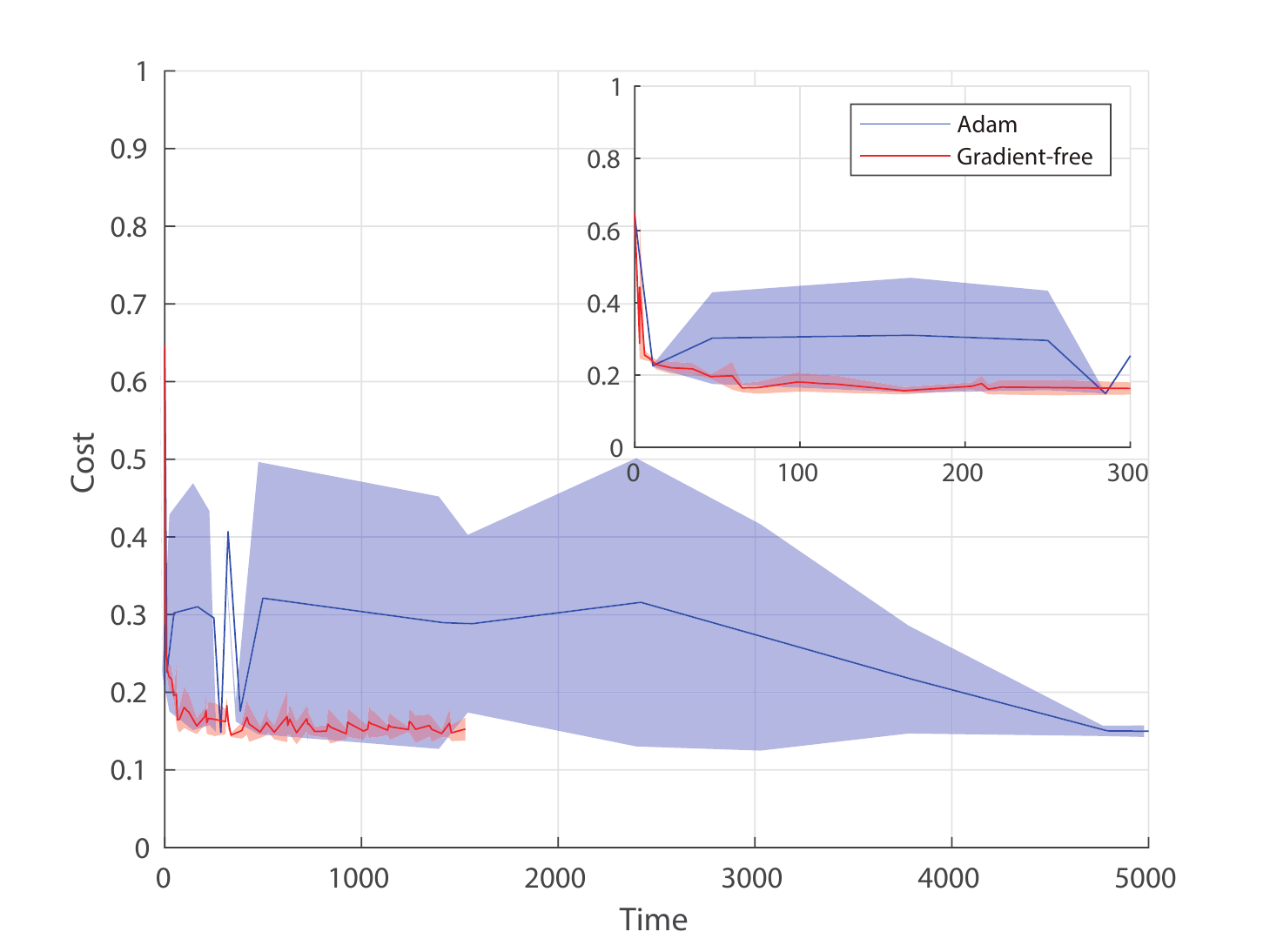}}
	\subfigure[] {\includegraphics[width=.45\textwidth]{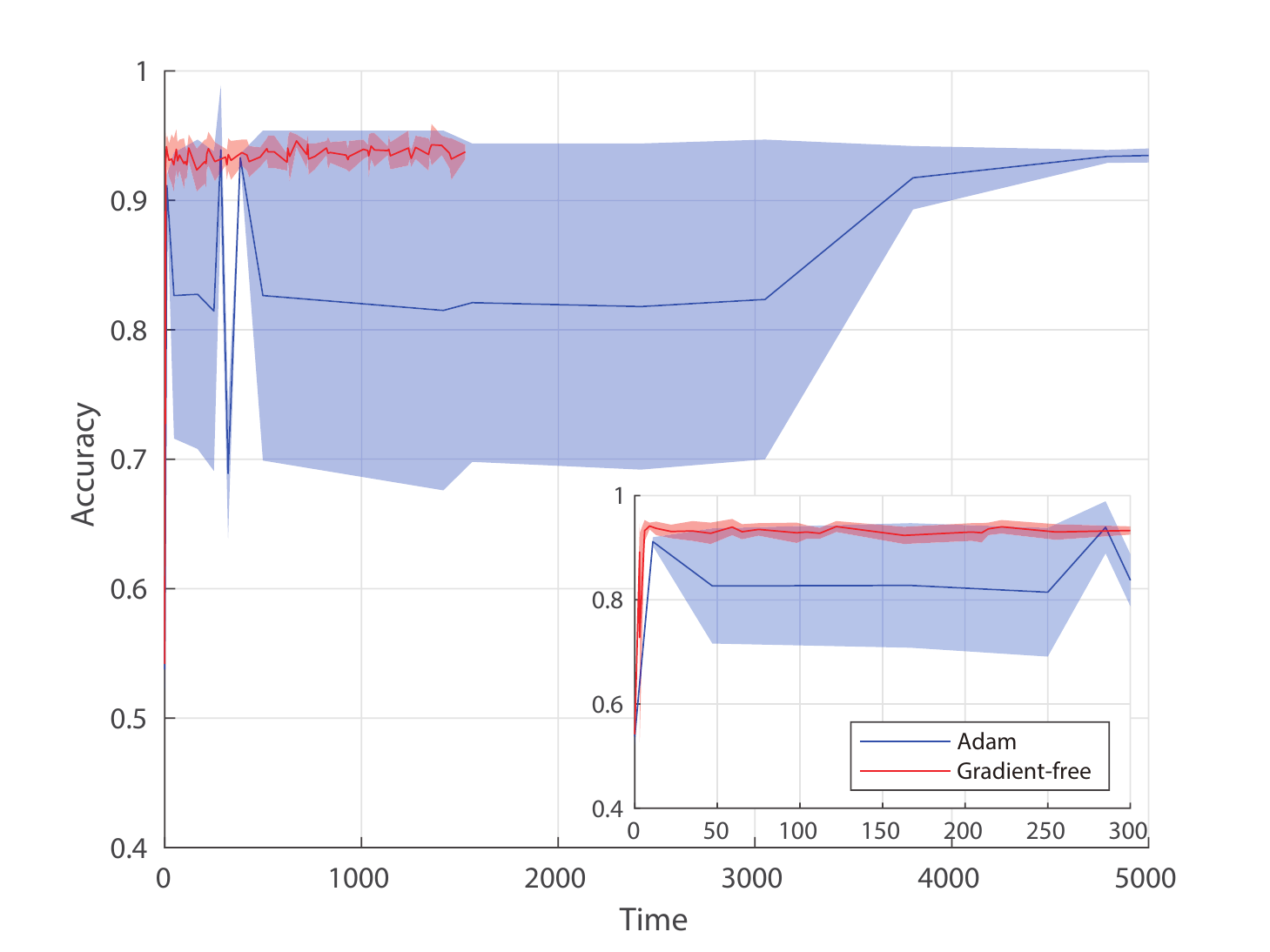}}
	\caption{The classification accuracy and the cost function against the time of 15 loops for the two-class classification task with bit flip noise in quantum gate, where (a) is the classification accuracy and (b) is the cost function.}
	\label{Fig:3}
\end{figure*}

We first consider the case that the quantum gate without noise. Fig.\ref{Fig:2} shows the classification accuracy(a) and the value of cost(b) against time for a two-class classification task. As it's shown in Fig.\ref{Fig:2}(a), in the beginning of optimizing, the fluctuation range of GFO algorithm is larger than that of Adam optimizer when the value of cost decreases. However, the convergence of GFO algorithm tends to be stable, appearing to be an advantage over Adam optimizer. As shown in Fig.\ref{Fig:2}(b), GFO algorithm can achieve a high classification accuracy faster than Adam optimizer.

Furthermore, we consider the quantum gate which is affected by five different types of quantum noise, i.e., bit flip, depolarizing, phase damping, phase flip, and amplitude damping. We set that the rotation gate in quantum device has one of the five types of noise with 5$\%$ probability. 

Results in Fig.\ref{Fig:3} show the case of bit flip noise. Fig.\ref{Fig:3}(a) demonstrates that the convergence ability of GFO algorithm is better than Adam optimizer. While from Fig.\ref{Fig:3}(b), one can see that GFO algorithm is able to reach a high accuracy faster than Adam optimizer.

\begin{figure*}[htbp]
	\centering
	\subfigure[] {\includegraphics[width=.45\textwidth]{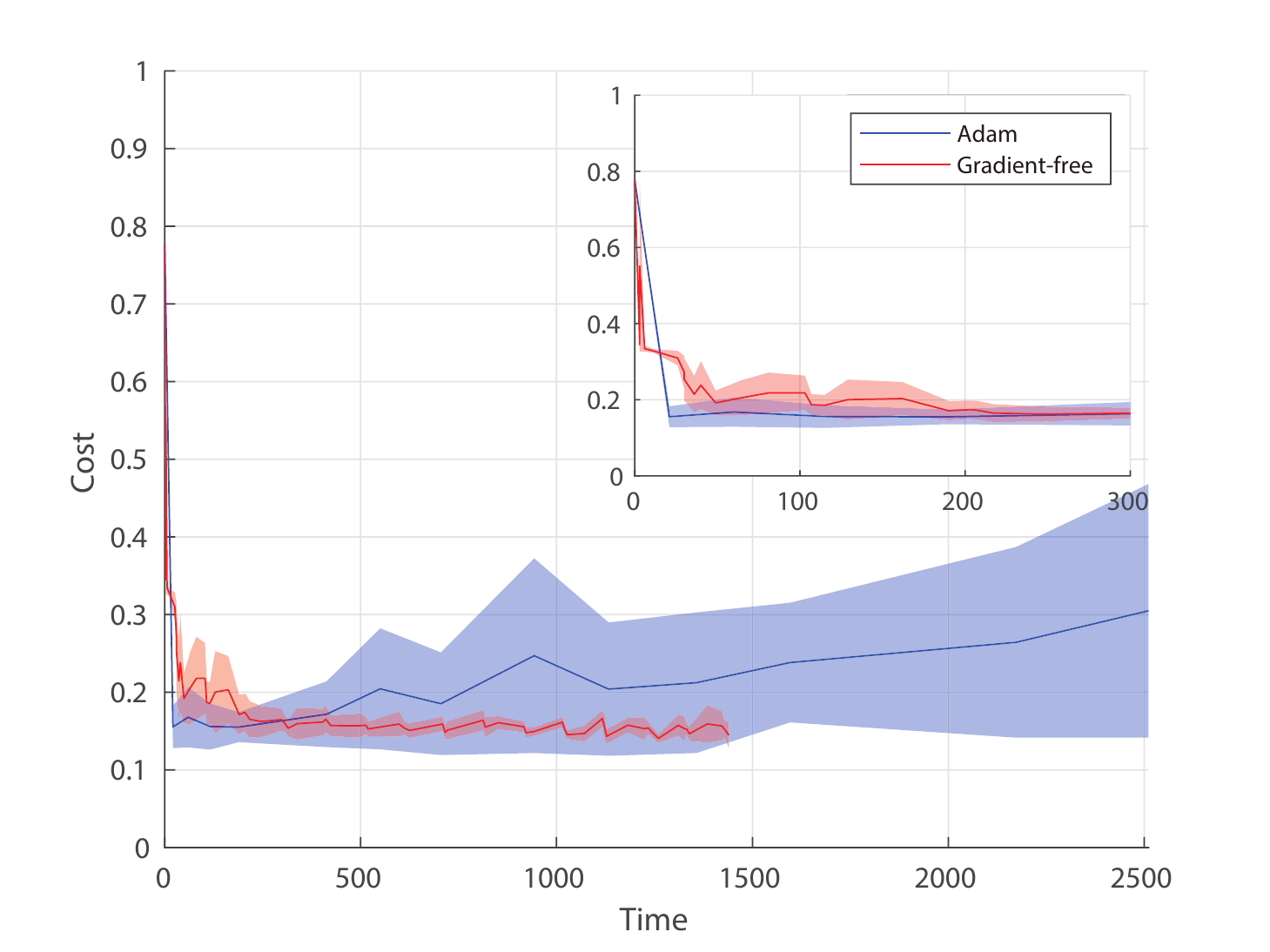}}
	\subfigure[] {\includegraphics[width=.45\textwidth]{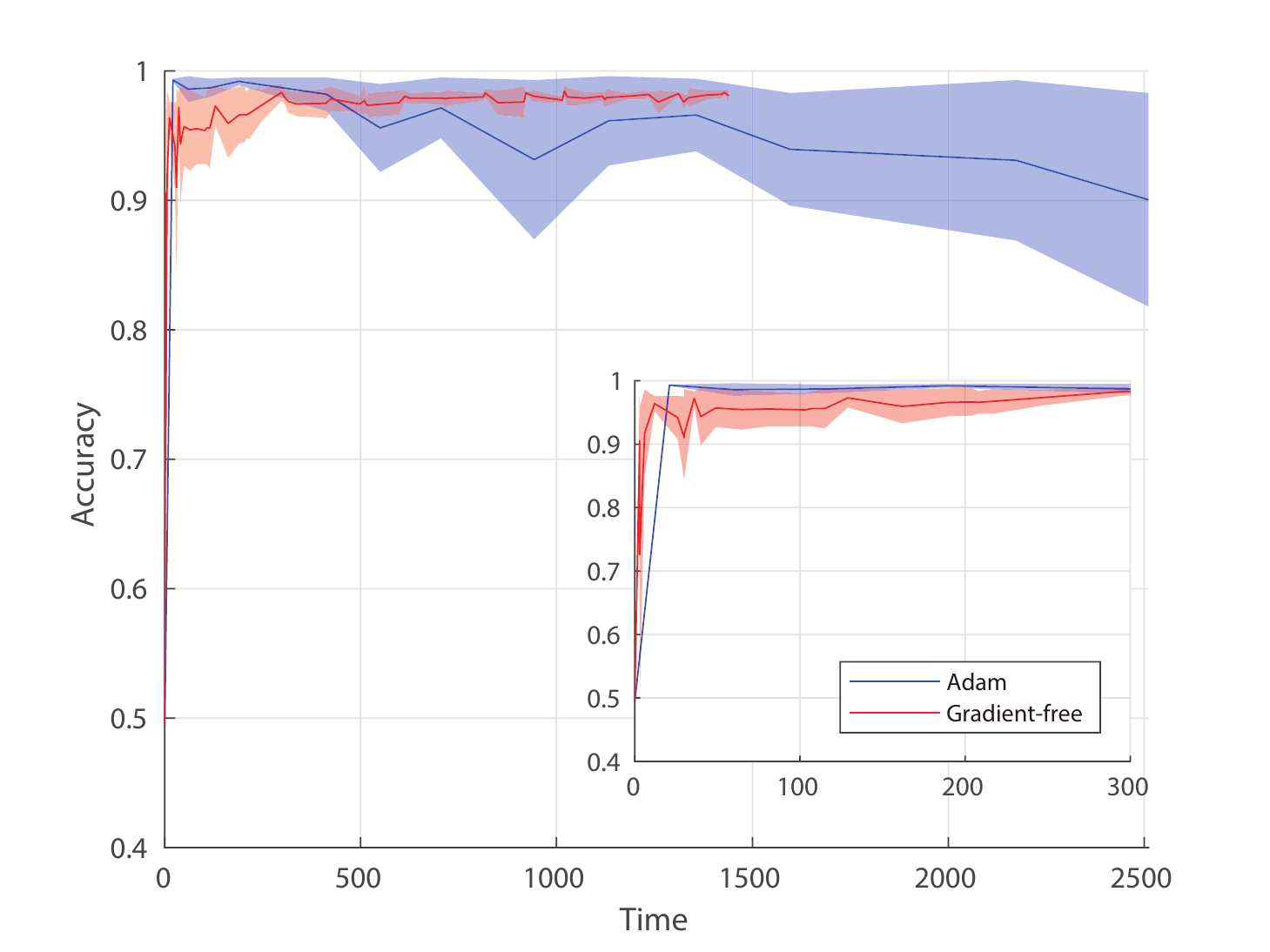}}
	\caption{The classification accuracy and the cost value against the time of 15 loops for the two-class classification task with depolarizing noise in quantum gate, where (a) is the classification accuracy and (b) is the cost value.}
	\label{Fig:4}
\end{figure*}

Results in Fig.\ref{Fig:4} show the case of depolarizing noise. Fig.\ref{Fig:4}(a) shows that the convergence ability of GFO algorithm is a bit weaker than that of Adam optimizer in the early stage but better later. Fig.\ref{Fig:4}(b) shows that GFO algorithm can achieve a high classification accuracy faster, but the Adam optimizer can reach a higher accuracy than the GFO algorithm eventually. As the training process goes on, the GFO algorithm can maintain a stable and high classification accuracy, which is better than Adam optimizer.

\begin{figure*}[htbp]
	\centering
	\subfigure[] {\includegraphics[width=.45\textwidth]{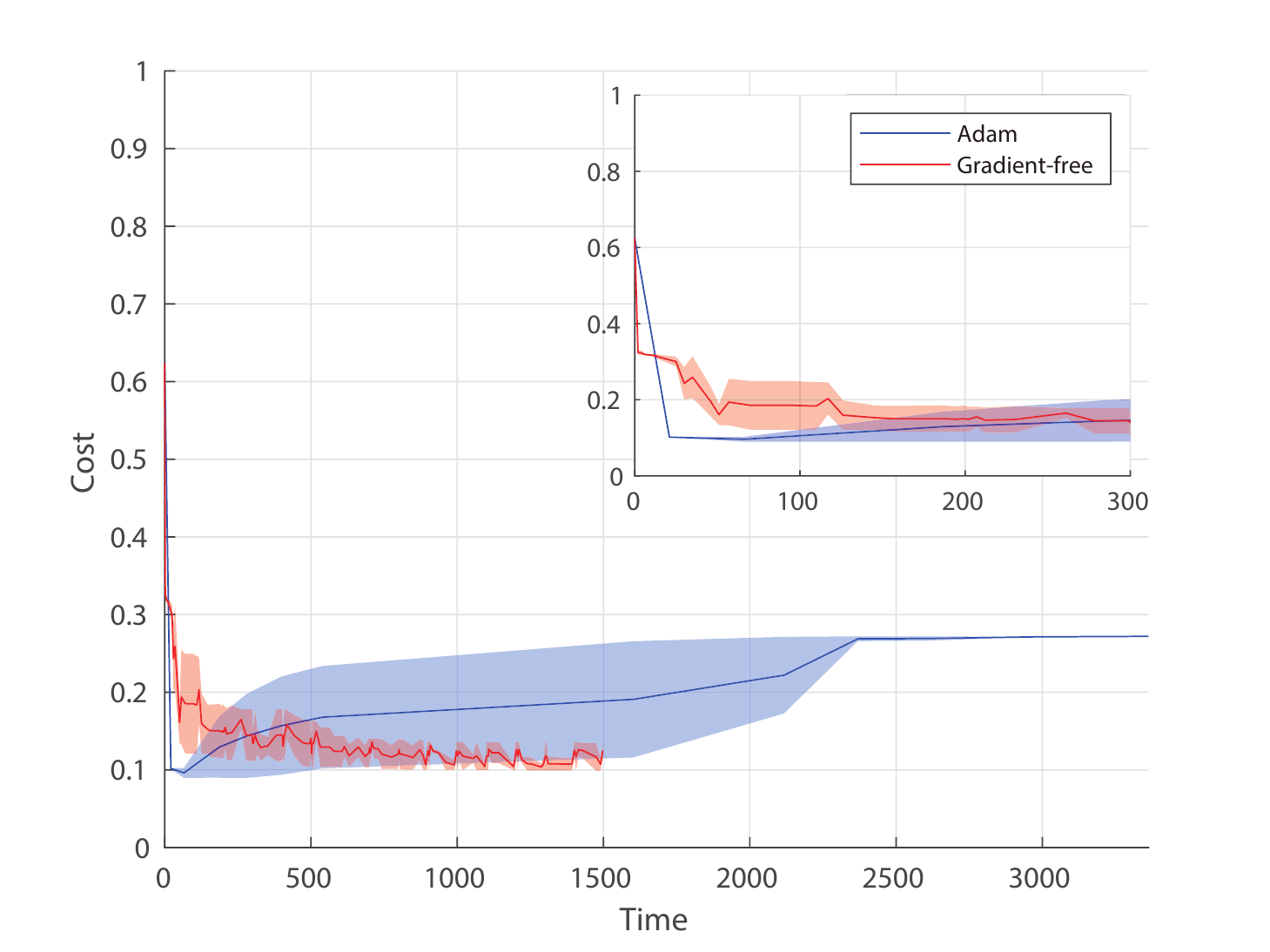}}
	\subfigure[] {\includegraphics[width=.45\textwidth]{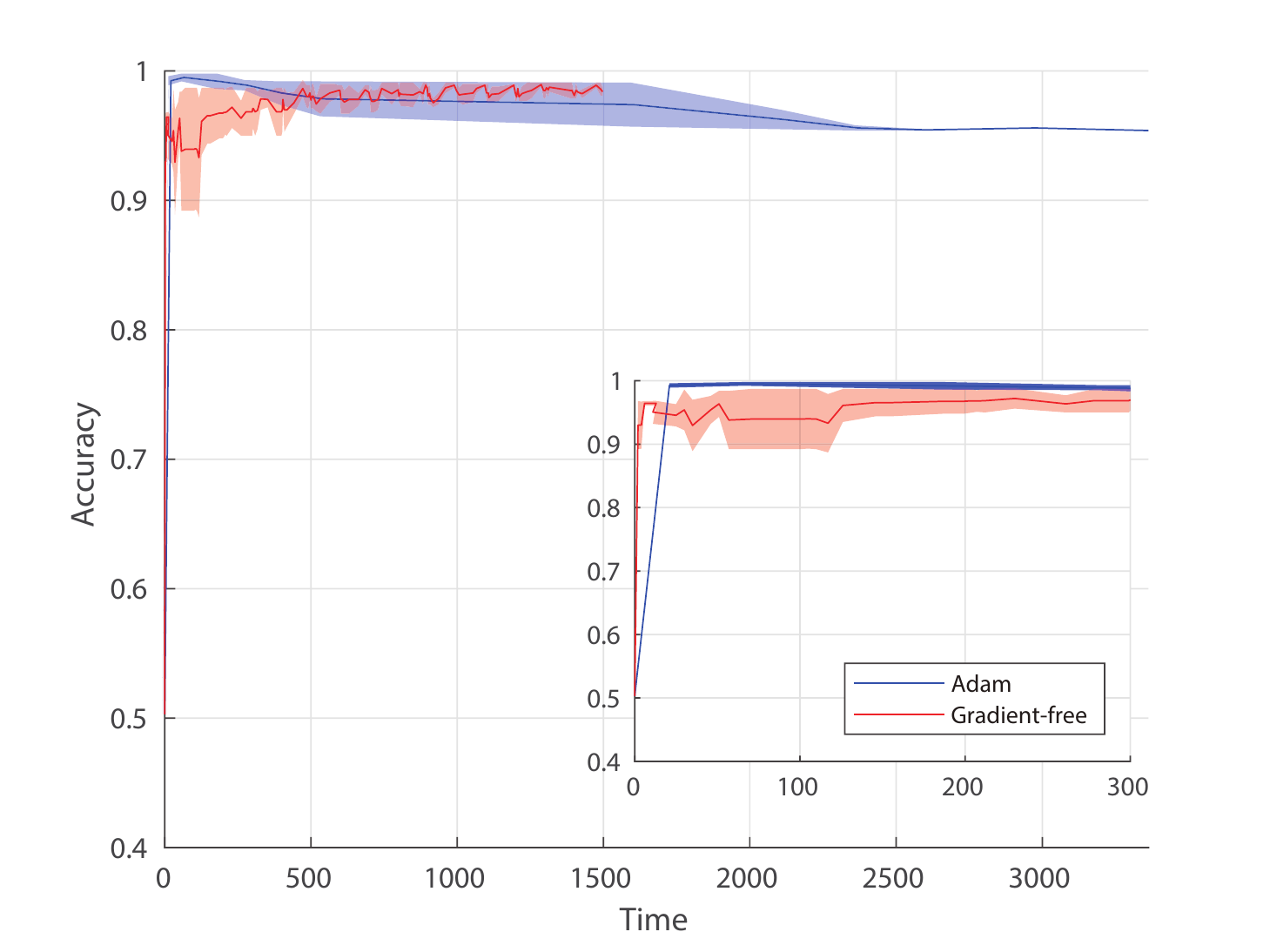}}
	\caption{The classification accuracy and the cost function against the time of 15 loops for the two-class classification task with phase damping noise in quantum gate, where (a) is the classification accuracy and (b) is the cost function.}
	\label{Fig:5}
\end{figure*}

\begin{figure}[!htbp]
	\centering
	\subfigure[] {\includegraphics[width=.45\textwidth]{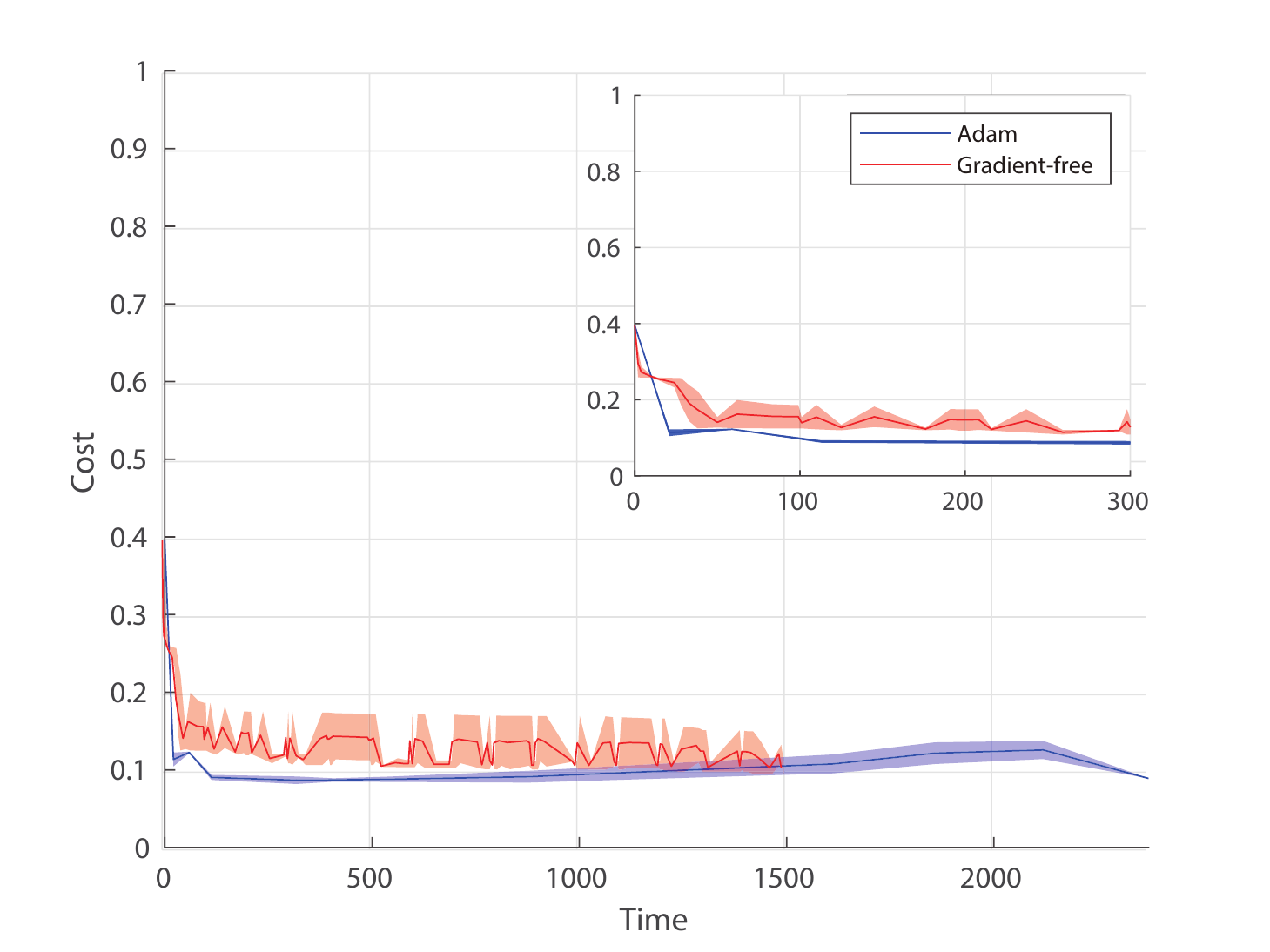}}
	\subfigure[] {\includegraphics[width=.45\textwidth]{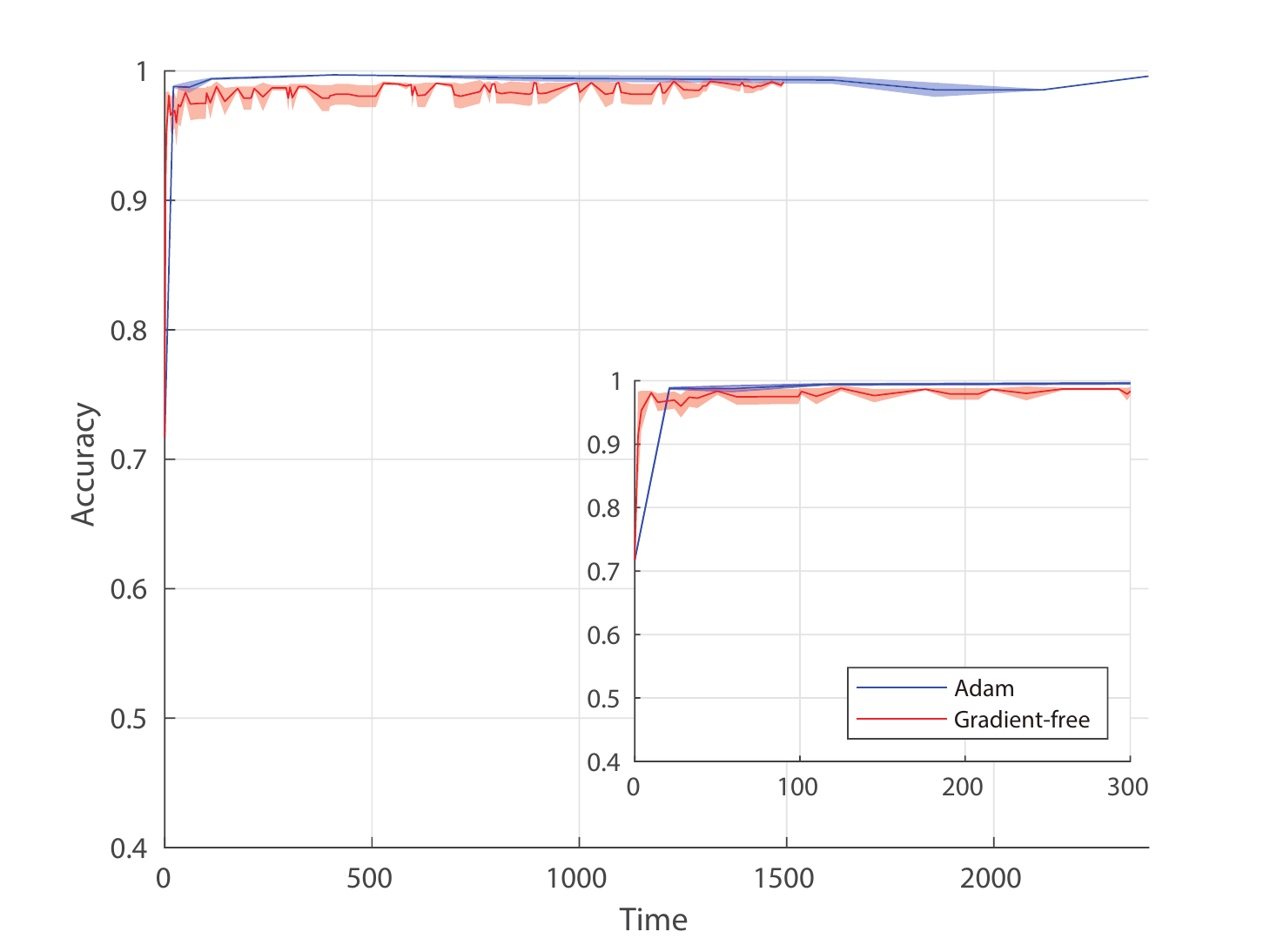}}
	\caption{The classification accuracy and the cost function against the time of 15 loops for the two-class classification task with phase flip noise in quantum gate, where (a) is the classification accuracy and (b) is the cost function.}
	\label{Fig:6}
\end{figure}

\begin{figure*}[h]
	\centering
	\subfigure[] {\includegraphics[width=.45\textwidth]{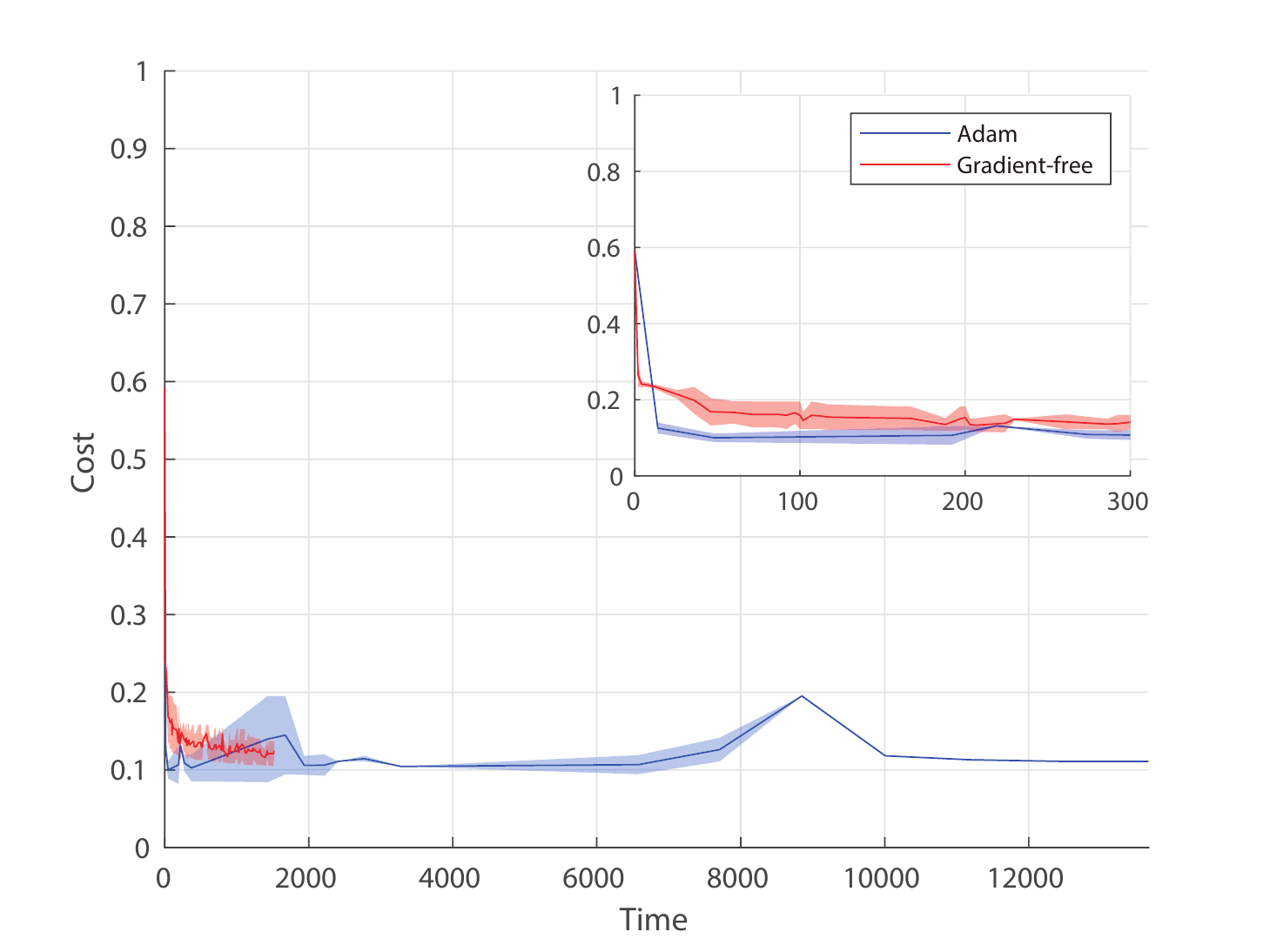}}
	\subfigure[] {\includegraphics[width=.45\textwidth]{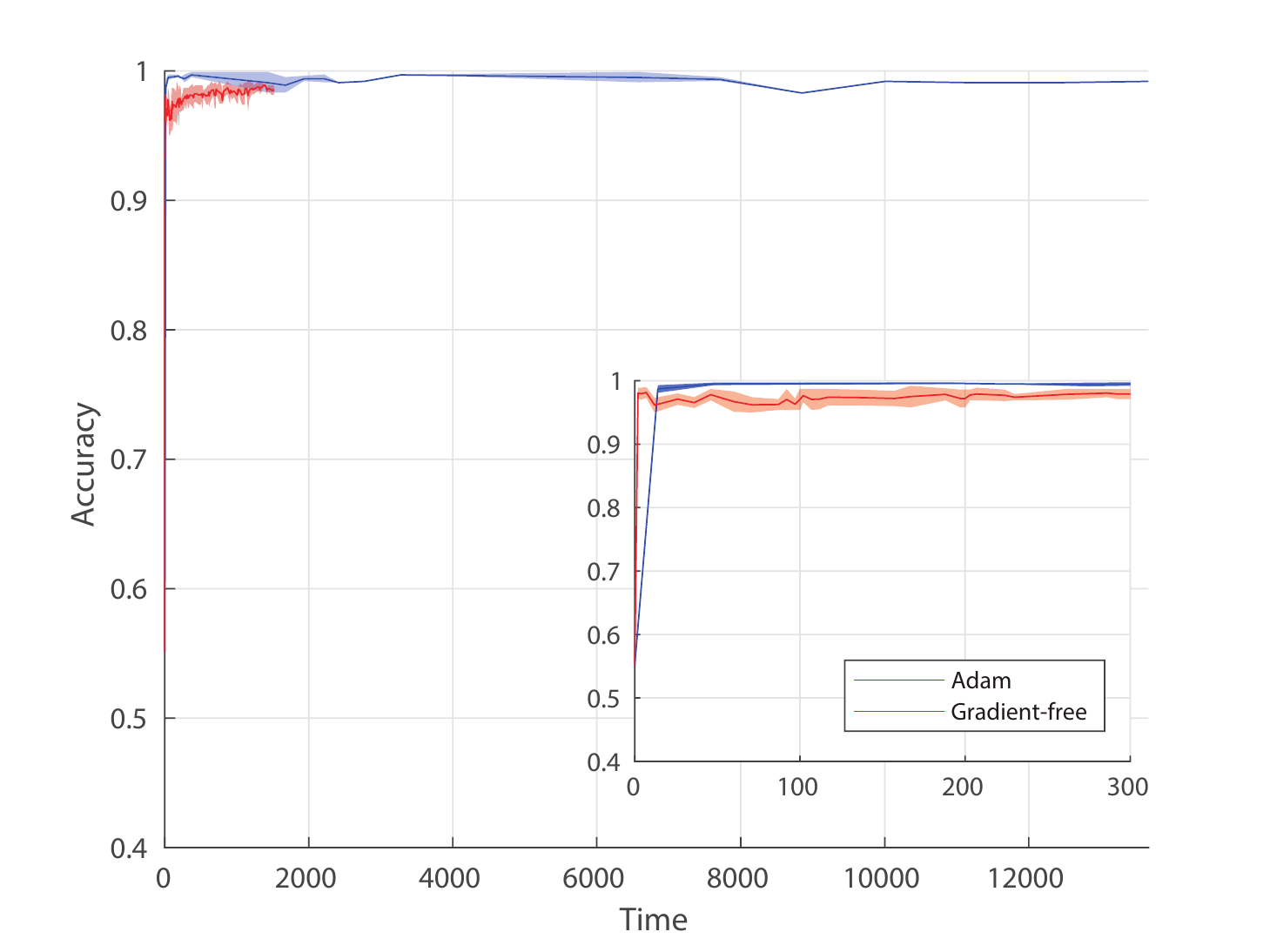}}
	\caption{The classification accuracy and the cost function against the time of 15 loops for the two-class classification task with amplitude damping noise in quantum gate, where (a) is the classification accuracy and (b) is the cost function.}
	\label{Fig:7}
\end{figure*}

Results in Fig.\ref{Fig:5} show the case of phase damping noise. Fig.\ref{Fig:5}(a) shows that GFO algorithm has ability of resisting the phase damping noise and keeping stable continuous convergence, while Adam optimizer has a better convergence ability than GFO algorithm but fluctuate during later time. Fig.\ref{Fig:5}(b) shows that GFO algorithm can achieve a high classification accuracy faster but with a large range of fluctuation in early time. Finally, GFO algorithm can maintain a stable and high classification accuracy, outperforming itself than Adam optimizer.

Results in Fig.\ref{Fig:6} show the case of phase flip noise. Fig.\ref{Fig:6}(a) shows that GFO algorithm does not has a good ability of resisting the phase flip noise as Adam optimizer. Fig.\ref{Fig:6}(b) shows that GFO algorithm can achieve a high classification accuracy more quick, but the eventual accuracy of Adam optimizer is higher.

Results in Fig.\ref{Fig:7} show the case of amplitude damping noise. As is shown in Fig.\ref{Fig:7}(a), the convergence ability of GFO algorithm is weaker than Adam optimizer. Fig.\ref{Fig:7}(b) shows that GFO algorithm can achieve a high classification accuracy faster but Adam optimizer can reach a higher accuracy eventually.

In general, GFO algorithm has a certain ability of resistance to noises and can keep the ability of continuous convergence. Although GFO algorithm is affected by the randomly initialized parameters, which makes the optimization performance of GFO algorithm fluctuate in early stage of training processes, GFO algorithm can always achieve a high classification accuracy faster than Adam optimizer and maintain stable in later stage. As gradients do not need to be calculated, GFO algorithm saves lots time of calculating and can quickly complete the training process of the single-qubit quantum classifier. While Adam optimizer is affected badly with different noise and use different time to finish the training processes in different noise environment.

\section{Discussion and conclusion}
In this work, we have proposed a gradient-free algorithm for parameter optimization in a single-qubit quantum classifier. By loading training data and parameters in a summation form, only one rotation gate have been used in quantum circuit. The proposed gradient-free algorithm has used to optimize only one parameter each time, resulting in a shorter time for completing the training process of the single-qubit classifier. Furthermore, we have simulated the classification tasks with the GFO single-qubit quantum classifier, and have compared the results with that using  Adam optimizer. The simulation results have shown that the single-qubit quantum classifier with proposed gradient-free optimization algorithm has obtained a higher accuracy and a shorter time than that using Adam optimizer.  Additionally, the single-qubit quantum classifier with proposed gradient-free optimization algorithm  has have a good performance in a noisy environment.

For the further work, we will focus on how to increase the the expression power of quantum circuit where parameters can optimized by this algorithm for other complexity classification tasks.

\section*{Acknowledgement}
This work is supported by the National Natural Science Foundation of China (61871234), and Postgraduate Research $\& $ Practice Innovation Program of Jiangsu Province (Grant KYCX19\_0900).


\begin{thebibliography}{20}

\bibitem{JBiamonte}J. Biamonte, P. Wittek, N. Pancotti, P. Rebentrost, N. Wiebe and S. Lloyd, Quantum machine learning, Nature 549, 195–202 (2017)

\bibitem{MBenedetti}M. Benedetti, E. Lloyd, S. Sack and M. Fiorentini, Parameterized quantum circuits as machine learning models, Quantum Science and Technology 4, 4 (2019)

\bibitem{SLTeresa}S. L. Teresa, R. R. Juan, and Z. David, Quantum kernels to learn the phases of quantum matter, Phys. Rev. A 105, 042432 (2022)

\bibitem{KMNaoko}K. M. Naoko and M. Kei, Fast and scalable classical machine-learning algorithm with similar performance to quantum circuit learning, Phys. Rev. A 104, 062411 (2021) 

\bibitem{KHWan}K. H. Wan, O. Dahlsten, H. Kristj\'{\ a}nsson, R. Gardner and M. S. Kim, Quantum generalisation of feedforward neural networks, npj Quantum Information 3, 36 (2017)

\bibitem{ETorrontegui}E. Torrontegui, J. J. Garcia-Ripoll, Unitary quantum perception as efficient universal approximator, Europhysics Letters 125 ,30004(2019).

\bibitem{JMArrazola}N. Killoran, T. R. Bromley, J. M. Arrazola, M. Schuld, N. Quesada and S. Lloyd, Continuous-variable quantum neural networks, Phys. Rev. Research 1, 033063 (2019)

\bibitem{AMari}A. Mari, T. R. Bromley, J. Izaac, M. Schuld, and N. Killoran, Transfer learning in hybrid classical-quantum neural networks, Quantum 4, 340 (2020) 

\bibitem{NKilloran}M. Schuld, N. Killoran, Quantum machine learning in feature Hilbert spaces, Phys. Rev. Lett. 122, 040504(2019) 

\bibitem{AGilyn}A. Gily\'{\ e}n, S. Arunachalam and N. Wiebe. Optimizing quantum optimization algorithms via faster quantum gradient computation. Proceedings of the Thirtieth Annual ACM-SIAM Symposium on Discrete Algorithms. 1425-1444(2019)

\bibitem{ECampos}E. Campos, A. Nasrallah and J. Biamonte, Abrupt transitions in variational quantum circuit training, Phys. Rev. A 103, 032607 (2021)

\bibitem{NWiebe}M. Schuld, A. Bocharov, K. M. Svore and N. Wiebe, Circuit-centric quantum classifiers, Phys. Rev. A 101, 032308 (2020)

\bibitem{SDangwal}S. Adhikary, S. Dangwal and D. Bhowmik, Supervised learning with a quantum classifier using multi-level systems, QUANTUM INF PROCESS 19, 89(2020) 

\bibitem{RKune}A. Chalumuri, R. Kune and B. S. Manoj, A hybrid classical-quantum approach for multi-class classification, QUANTUM INF PROCESS 20, 119(2021) 

\bibitem{AChalumuri}A. Chalumuri, R. Kune ,B. S. Manoj, C. W. Hsing and Y. Jer. Kao, An end-to-end trainable hybrid classical-quantum classifier, arXiv:2102.02416(2021) 

\bibitem{ASBhatia}A. S. Bhatia, M. K. Saggi, A. Kumar and S. Jain, Matrix product state–based quantum classifier, Neural Computation, 31, 1499-1517(2019) 

\bibitem{APerezSalinas}P. S. Adri\'{\ a}n, C. L. Alba, G. F. Elies , and I. L. Jos\'{\ e}, Data re-uploading for a universal quantum classifier, Quantum 4, 226(2020)

\bibitem{SAdhikary}S. Adhikary, An entanglement enhanced training algorithm for supervised quantum classifiers, arXiv:2006.13302(2020)

\bibitem{AQZhangMulti}A. Q. Zhang, X. Y. He and S. M. Zhao, Quantum algorithm for neural network enhanced multi-class parallel classification, arXiv:2203.04097(2022)

\bibitem{ZHolmes}Z. Holmes, K. Sharma, M. Cerezo and P. J. Coles, Connecting ansatz expressibility to gradient magnitudes and barren plateaus, arXiv:2101.02138 (2021).

\bibitem{ASkolik}A. Skolik, J. R. McClean, M. Mohseni, P. van der Smagt and M. Leib, Layerwise learning for quantum neural networks. Quantum Machine Intelligence 3, 5 (2021)

\bibitem{GIannelli}G. Iannelli, K. Jansen, Noisy Bayesian optimization for variational quantum eigensolvers. arXiv: 2112.00426v1

\bibitem{MOstaszewski}M. Ostaszewski, E. Grant and M. Benedetti, Structure optimization for parameterized quantum circuits, Quantum 5, 391 (2019)

\bibitem{PaoloC}P. Comelli, P. Ferragina, M.N. Granieri and F. Stabile, Optical recognition of motor vehicle license plates, IEEE Transactions on Vehicular Technology 44, 790-799(1995)

\bibitem{VilleB}V. Bergholm, J. Izaac, M. Schuld, C. Gogolin, M. S. Alam, S. Ahmed, J. M. Arrazola, C. Blank, A. Delgado, S. Jahangiri, K. McKiernan, J. J. Meyer, Z. N., A. Sz\'{\ a}va, N. Killoran, Pennylane: Automatic differentiation of hybrid quantum-classical computations, arXiv:1811.04968(2018)
\end{thebibliography}
\end{document}